\documentclass[preprint,prl,showpacs,aps,amsmath,amssymb]{revtex4}

\usepackage{graphicx}

\begin{document}
\title{On Field Induced Diaelastic Effect in a Small Josephson Contact}

\author{S. Sergeenkov and F.M. Araujo-Moreira}
\affiliation{Grupo de Materiais e Dispositivos, Departamento de
F\'{i}sica, Universidade Federal de S\~{a}o Carlos, 13565-905
S\~{a}o Carlos, SP, Brazil }


\date{\today}

\begin{abstract}

An analog of the diaelastic effect is predicted to occur in a
small Josephson contact with Josephson vortices manifesting itself
as magnetic field induced softening of the contact shear modulus
$C(T,H)$. In addition to Fraunhofer type {\it field} oscillations,
$C(T,H)$ is found to exhibit pronounced flux driven {\it
temperature} oscillations near $T_C$.
\end{abstract}


\maketitle

{\bf 1. Introduction.} Inspired by new possibilities offered by
the cutting-edge nanotechnologies,  the experimental and
theoretical physics of increasingly sophisticated mesoscopic
quantum devices heavily based on Josephson junctions (JJ) and
their arrays (JJA) is becoming one of the most exciting and
rapidly growing areas of modern science (for the recent reviews,
see, e.g.,~\cite{1,2,3,4} and further references therein ). In
particular, a remarkable increase of the measurements technique
resolution  made it possible to experimentally detect such
interesting phenomena as flux avalanches~\cite{5}, geometric
quantization~\cite{6}, flux driven oscillations of heat
capacity~\cite{7}, reentrant-like behavior~\cite{8}, manifestation
of $\pi$-contacts ~\cite{9}, $R$-$C$ crossover ~\cite{9a},
unusually strong coherent response~\cite{10}, Josephson analog of
the fishtail effect~\cite{11}, geometric resonance  and field
induced Kosterlitz-Thouless transition ~\cite{12}. Among the
numerous theoretical predictions (still awaiting their
experimental verification) one could mention electro- and
magnetostriction~\cite{13}, field induced polarization effects
~\cite{14}, analog of magnetoelectric effect~\cite{15}, nonlinear
Seebeck effect and thermal conductivity~\cite{16}, stress induced
effects~\cite{17}, chemomagnetism ~\cite{18}, magnetoinductance
effects ~\cite{19}, implications of dipolar interactions for
wireless connection between Josephson qubits ~\cite{20} and for
weakening of the Coulomb blockade~\cite{21}, proximity-induced
superconductivity in graphene ~\cite{21a} and anomalous Josephson
current in topological insulators ~\cite{21b}.

Turning to the subject of this Letter, let us recall that when an
elastic solid contains a region with compressibility different
from the bulk one, the applied stress $\sigma$ induces a spatially
inhomogeneous strain field $\epsilon$ around this region, which
results in the softening of its shear modulus $C$. This
phenomenon, known as diaelastic effect (DE), usually occurs in
materials with pronounced defect structure~\cite{22}. By
association, Josephson vortices can be considered as defects
related inclusions within tunneling contacts. Therefore, one could
expect an appearance of magnetic field induced analog of DE in
Josephson structures as well. By introducing an elastic response
of JJ to an effective stress field, in what follows we shall
discuss a possible manifestation of this novel interesting effect
in a small contact under an applied magnetic field.

{\bf 2. Model.} The temperature and field dependence of the
elastic shear modulus $C(T,H)$ of the Josephson structure can be
defined as follows (Cf. Ref.~\cite{17}):
\begin{equation}
\frac{1}{C(T,H)}=\left [\frac{d \epsilon (T,H,
\sigma)}{d\sigma}\right ]_{\sigma =0}
\end{equation}
where $\sigma$ is an applied stress and strain field $\epsilon$ in
the contact area is related to the stress dependent Josephson
critical current $I_C$ as follows ($V$ is the volume of the
sample)~\cite{17}:
\begin{equation}
\epsilon (T,H, \sigma)=\left (\frac{\Phi_0}{2\pi V}\right )
\frac{dI_C(T,H,\sigma)}{d\sigma}
\end{equation}
For simplicity and to avoid self-field effects, in what follows we
consider a small Josephson contact of length $w<\lambda _J$
($\lambda _J=\sqrt{\Phi _0/\mu _0dj_{c}}$ is the Josephson
penetration depth) placed in a strong enough magnetic field (which
is applied normally to the contact area) such that $H>\Phi _0/2\pi
\lambda _Jd$, where $d=2\lambda _{L}+t$, $\lambda _{L}$ is the
London penetration depth, and $t$ is an insulator thickness.

Recall that the critical current of such a contact in applied
magnetic field is governed  by a Fraunhofer-like
dependence~\cite{23}:
\begin{equation}
I_C(T,H,\sigma)=I_C(T,0,\sigma)\left [\frac{\sin \varphi
(T,H,\sigma ) }{\varphi (T,H,\sigma )}\right ]
\end{equation}
where $\varphi (T,H,\sigma )=\pi \Phi (T,H,\sigma )/\Phi _0$ with
$\Phi (T,H,\sigma )=Hwd(T,\sigma )$ being the temperature and
stress dependent flux through the contact area, and
$I_C(T,0,\sigma)\propto e^{-t/\xi}$ is the stress dependent
zero-field Josephson critical current with $\xi$ being a
characteristic (decaying) length and $t(\sigma)$ the stress
dependent thickness of the insulating layer (see below).

Notice that in non-zero applied magnetic field $H$, there are two
stress-induced contributions to the critical current $I_C$, both
related to decreasing of the insulator thickness under pressure.
First of all, it was experimentally observed~\cite{24} that the
tunneling dominated critical current of granular superconductors
exponentially increases under compressive stress $\sigma$, viz.
$I_C\propto e^{\kappa \sigma }$. More specifically, the critical
current at $\sigma =9 kbar$ was found to be three times higher its
value at $\sigma =1.5 kbar$, clearly indicating a
weak-links-mediated origin of the phenomenon. Hence, for small
enough $\sigma $ we can safely assume that~\cite{17} $t(\sigma
)\simeq t(0)(1-\beta \sigma)$. As a result, we have two
stress-induced effects in Josephson contacts: (a) amplitude
modulation leading to the explicit stress dependence of the
zero-field current
\begin{equation}
I_C(T,0,\sigma )=I_C(T,0,0)e^{\gamma \sigma}
\end{equation}
with $\gamma =\beta t(0)/\xi$, and (b) phase modulation leading to
the explicit stress dependence of the flux
\begin{equation}
\Phi (T,H,\sigma )=Hwd(T,\sigma )
\end{equation}
with
\begin{equation}
d(T,\sigma )=2\lambda _{L}(T)+t(0)(1-\beta \sigma)
\end{equation}
\begin{figure}
\includegraphics[width=10cm]{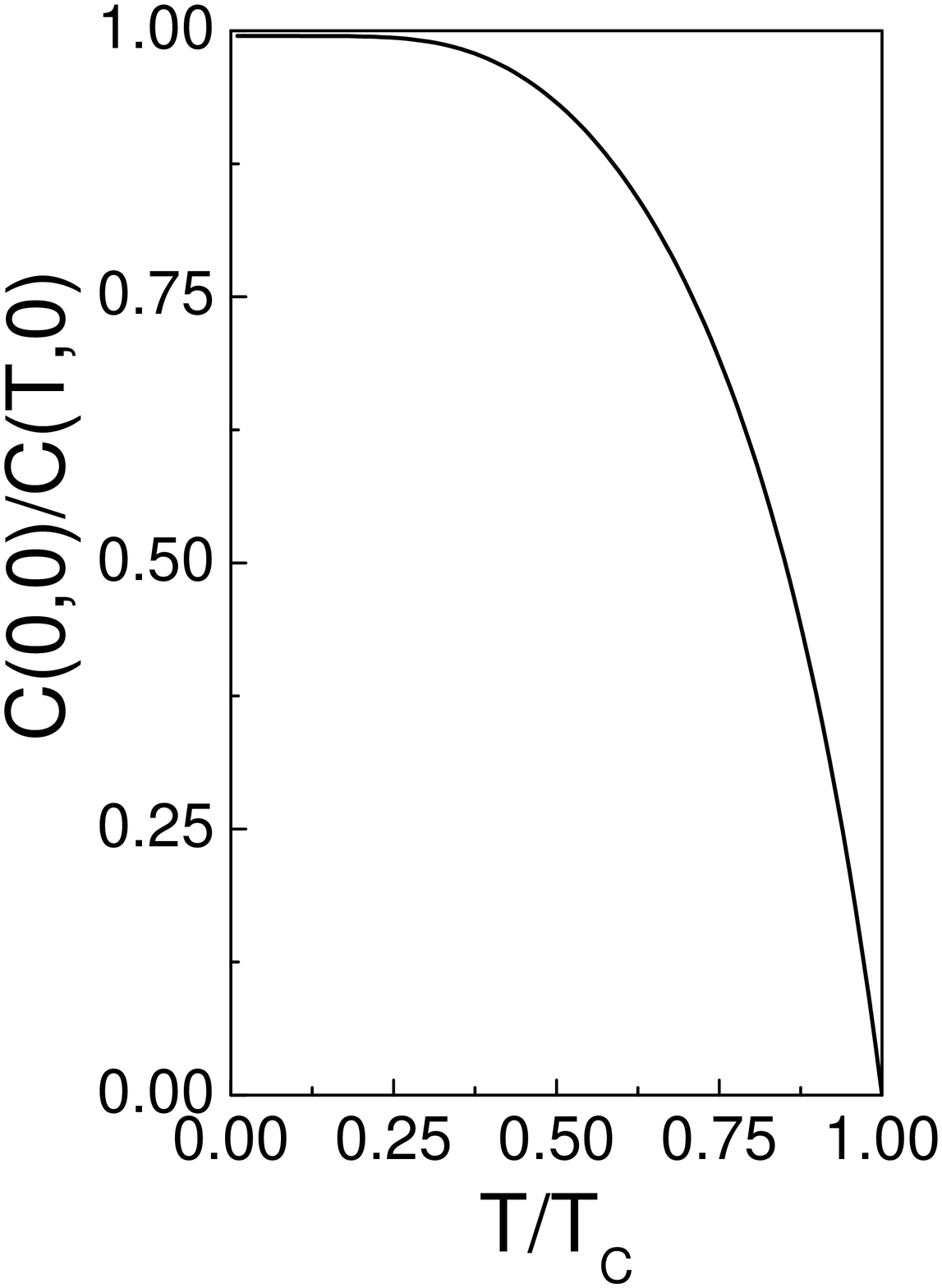}
\caption{Temperature dependence of the normalized inverse shear
modulus $C(0,0)/C(T,0)$ of a single short contact in zero magnetic
field according to Eqs.(1)-(12).} \label{fig:fig1}
\end{figure}
\begin{figure*}
\centerline{\includegraphics[width=15cm]{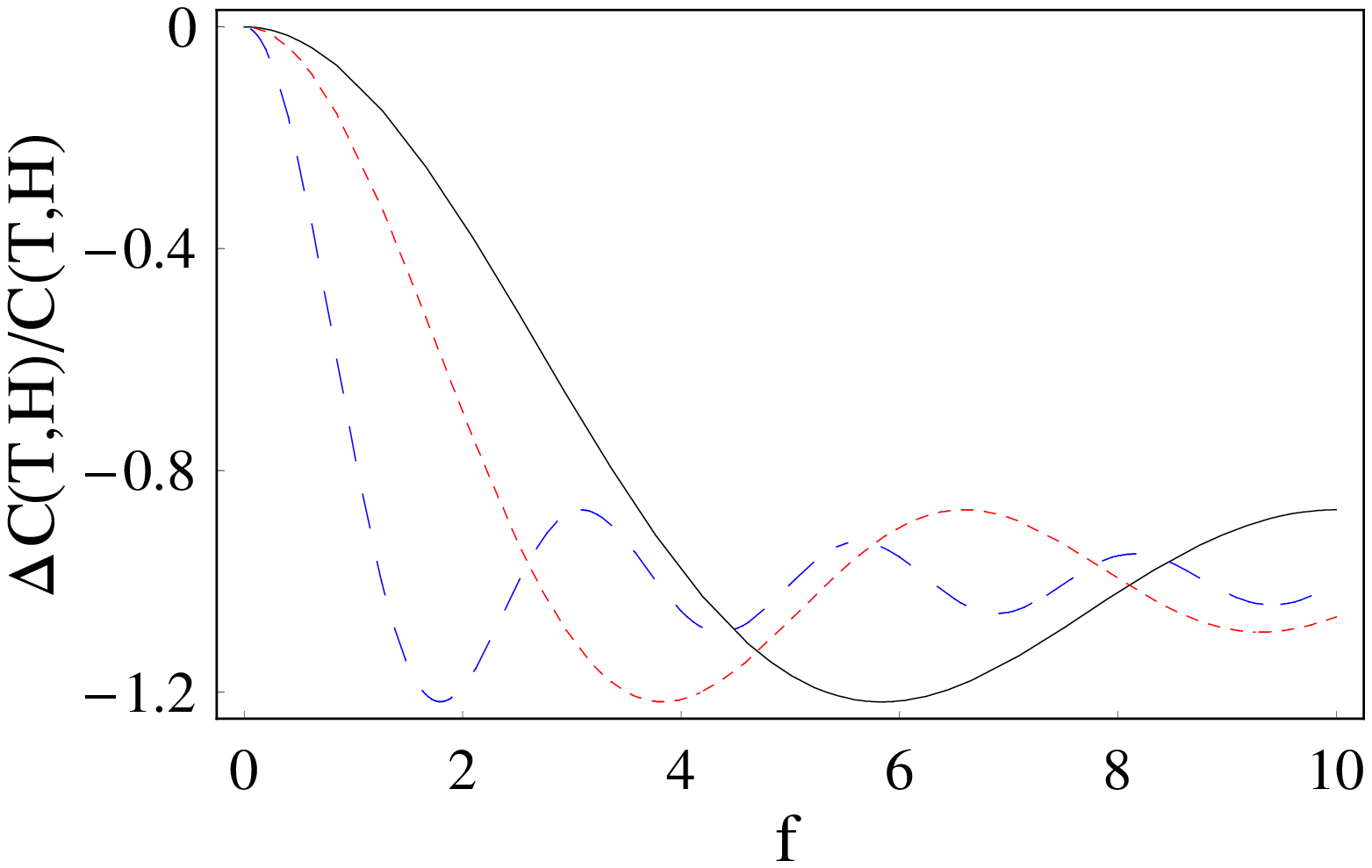}}\vspace{0.25cm}
\caption{ The field dependence of the diaelastic effect $\Delta
C(T,H)/C(T,H)$ for different temperatures: $T=0$ (solid line),
$T=0.5T_C$ (dotted line), and $T=0.9T_C$ (dashed line).}
\label{fig:fig2}
\end{figure*}
Finally, in view of Eqs.(1)-(6), the temperature and field
dependence of the small single junction shear modulus $C(T,H)$
reads:
\begin{equation}
\frac{1}{C(T,H)}=\frac{1}{C(T,0)}\left
[F(T,H)-\frac{\xi}{d(T,0)}\frac{dF(T,H)}{d\log{H}}\right ]
\end{equation}
where
\begin{equation}
F(T,H)=\left [\frac{\sin \varphi}{\varphi}+\frac{\xi}{d(T,0)}
\left (\frac{\sin \varphi}{\varphi}-\cos \varphi \right )\right ]
\end{equation}
with
\begin{equation}
\varphi(T,H)=\frac{\pi \Phi (T,H,0)}{\Phi _0}=\frac{H}{H_0(T)}
\end{equation}
and
\begin{equation}
\frac{1}{C(T,0)}=\left (\frac{\Phi _0 \gamma^2}{2\pi V}\right
)I_C(T)
\end{equation}
Here, $H_0(T)=\Phi _0/\pi wd(T,0)$ with $d(T,0)=2\lambda
_{L}(T)+t(0)$, and for convenience we used a simplified definition
$I_C(T,0,0)\equiv I_C(T)$ for zero-field and zero-stress critical
current.

For the explicit temperature dependence of $I_C(T)$ we use the
analytical approximation of the BCS gap parameter (valid for all
temperatures)~\cite{16}, $\Delta (T)=\Delta (0)\tanh
\left(2.2\sqrt{\frac{T_{C}-T}{T}}\right)$ with $\Delta
(0)=1.76k_BT_C$ which governs the temperature dependence of the
Josephson critical current
\begin{equation}
I_C(T)=I_C(0)\left[ \frac{\Delta (T)}{\Delta (0)}\right] \tanh
\left[ \frac{\Delta (T)}{2k_{B}T}\right]
\end{equation}
while the temperature dependence of the London penetration depth
is governed by the two-fluid model~\cite{25}:
\begin{equation}
\lambda _{L}(T)=\frac{\lambda _{L}(0)}{\sqrt{1-(T/T_C)^4}}
\end{equation}
\begin{figure*}
\begin{center}
\includegraphics[width=10.5cm]{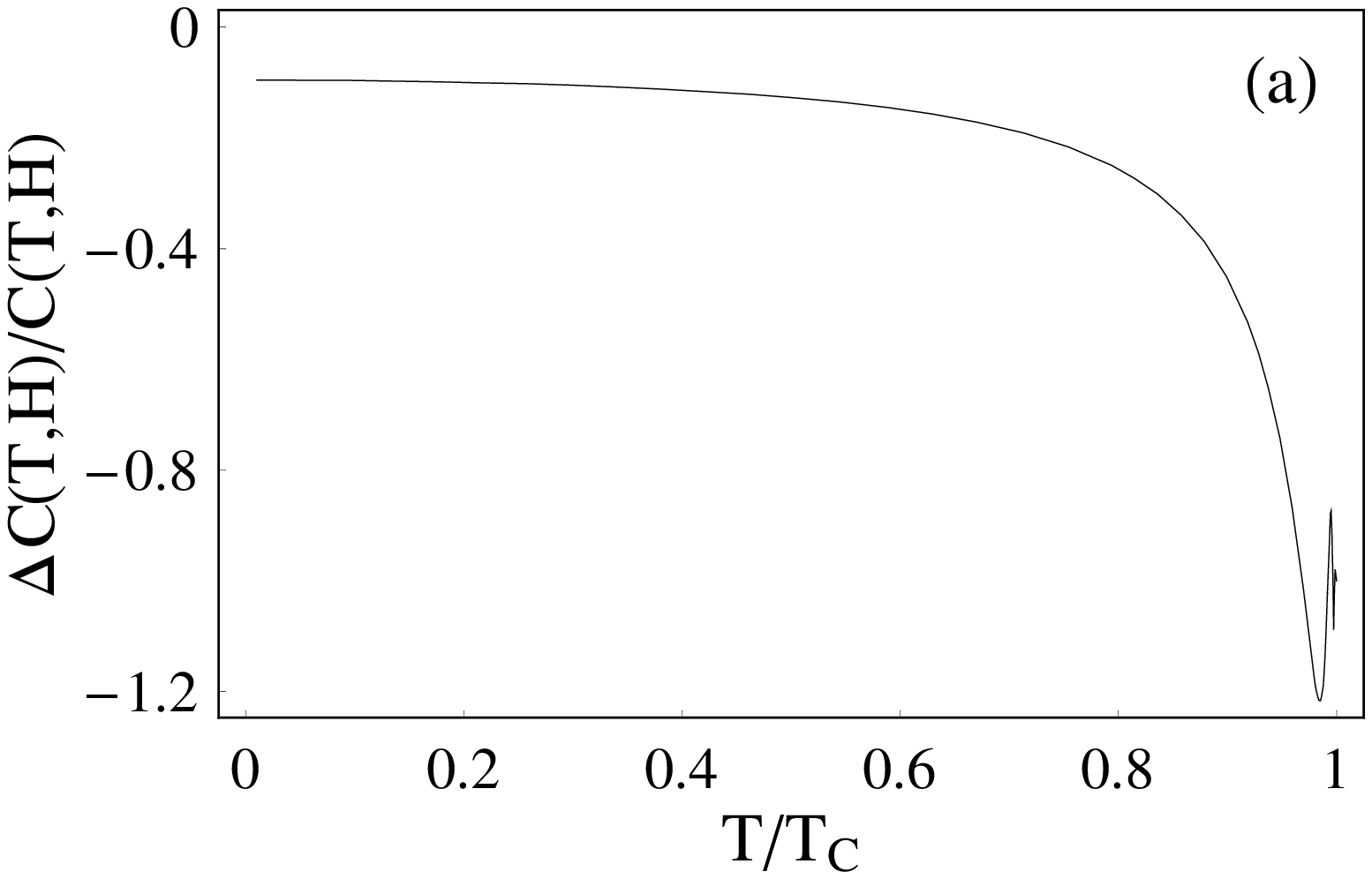}\vspace{0.0cm}
\includegraphics[width=10.5cm]{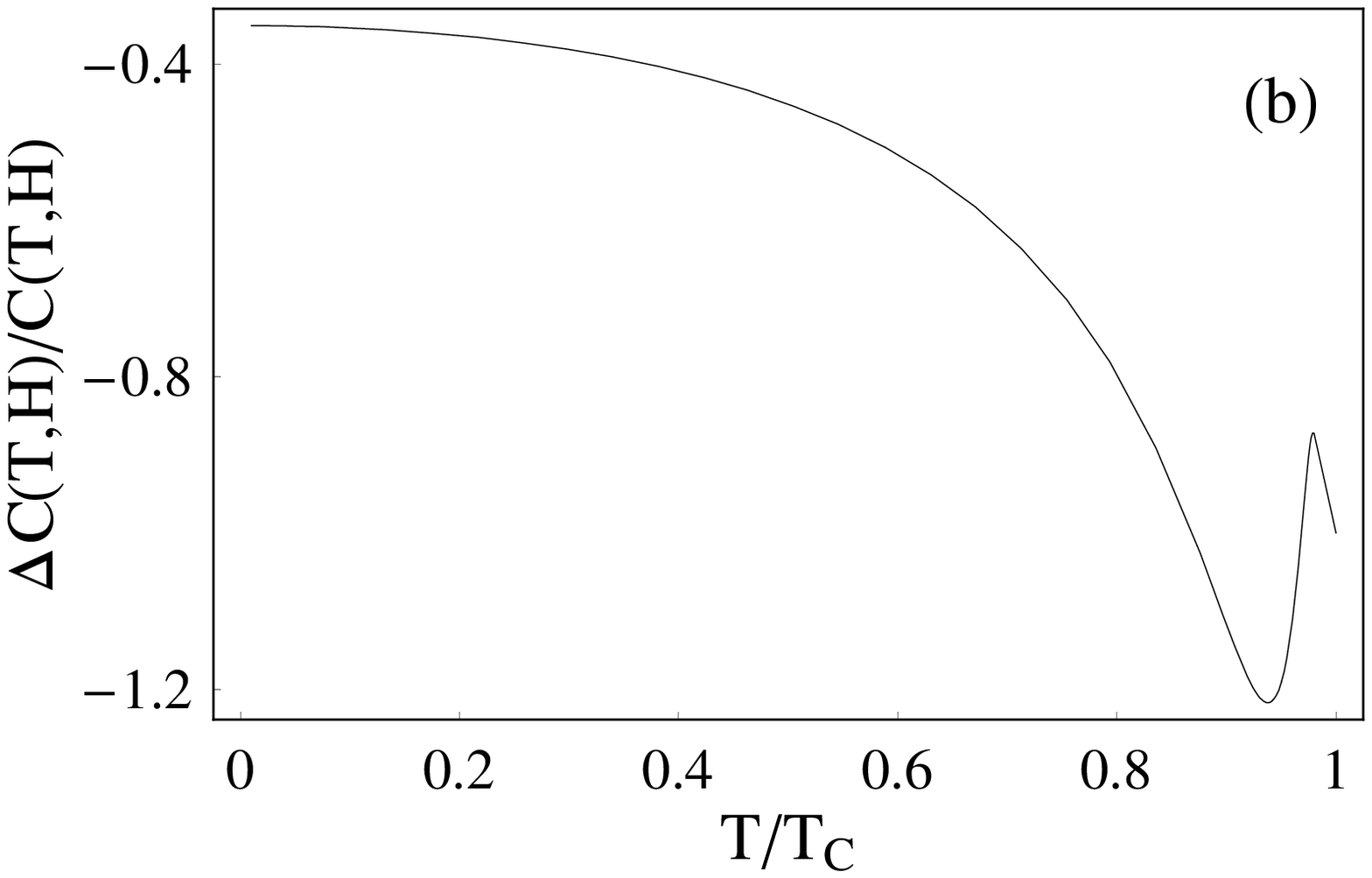}\vspace{0.0cm}
\includegraphics[width=10.5cm]{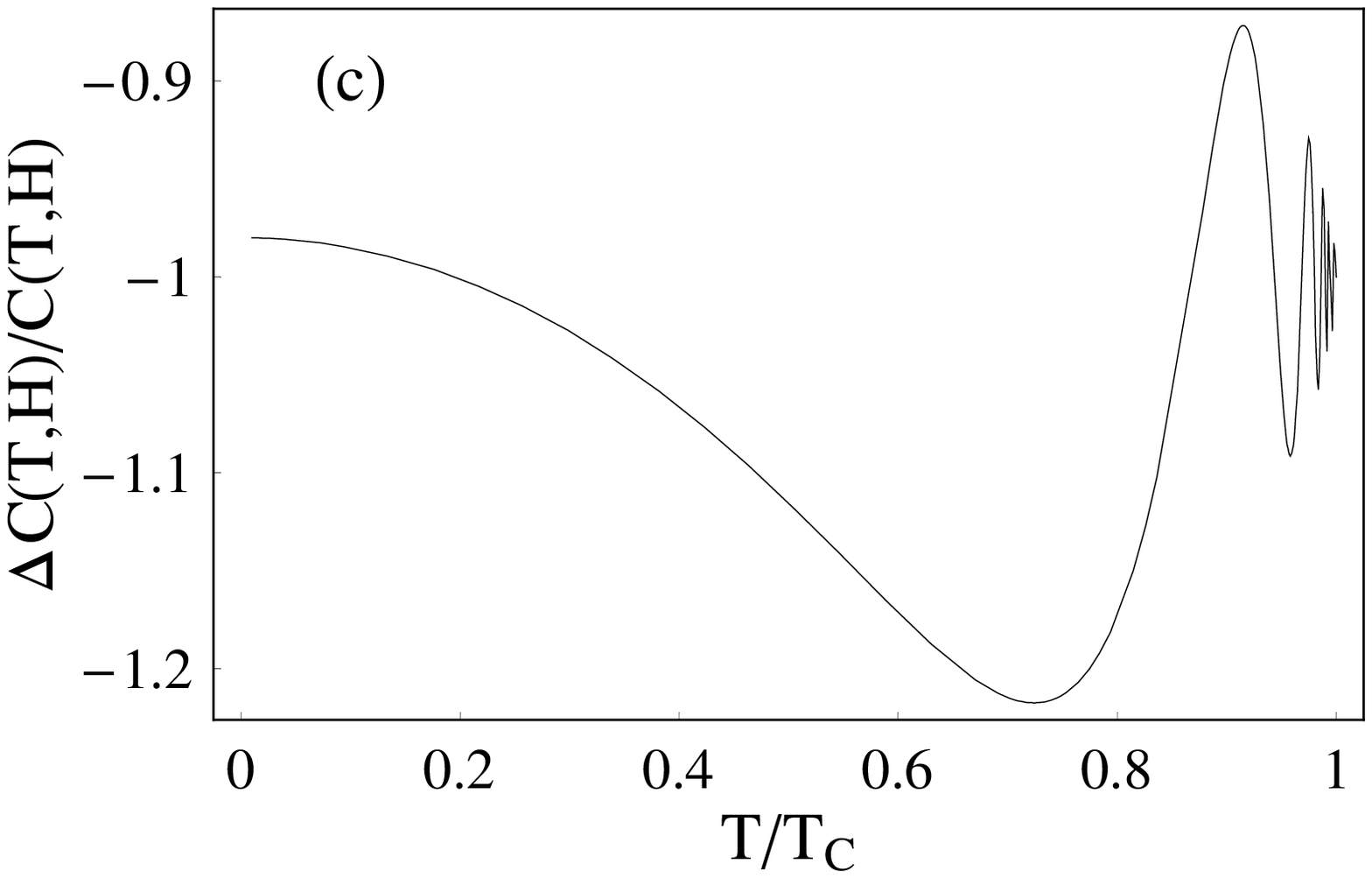}
\caption{Temperature dependence of the diaelastic effect $\Delta
C(T,H)/C(T,H)$ for different values of the frustration parameter
$f$: (a) $f=1$, (b) $f=3$, and (c) $f=5$.}\label{fig:fig3}
\end{center}
\end{figure*}

\begin{figure*}
\centerline{\includegraphics[width=15cm]{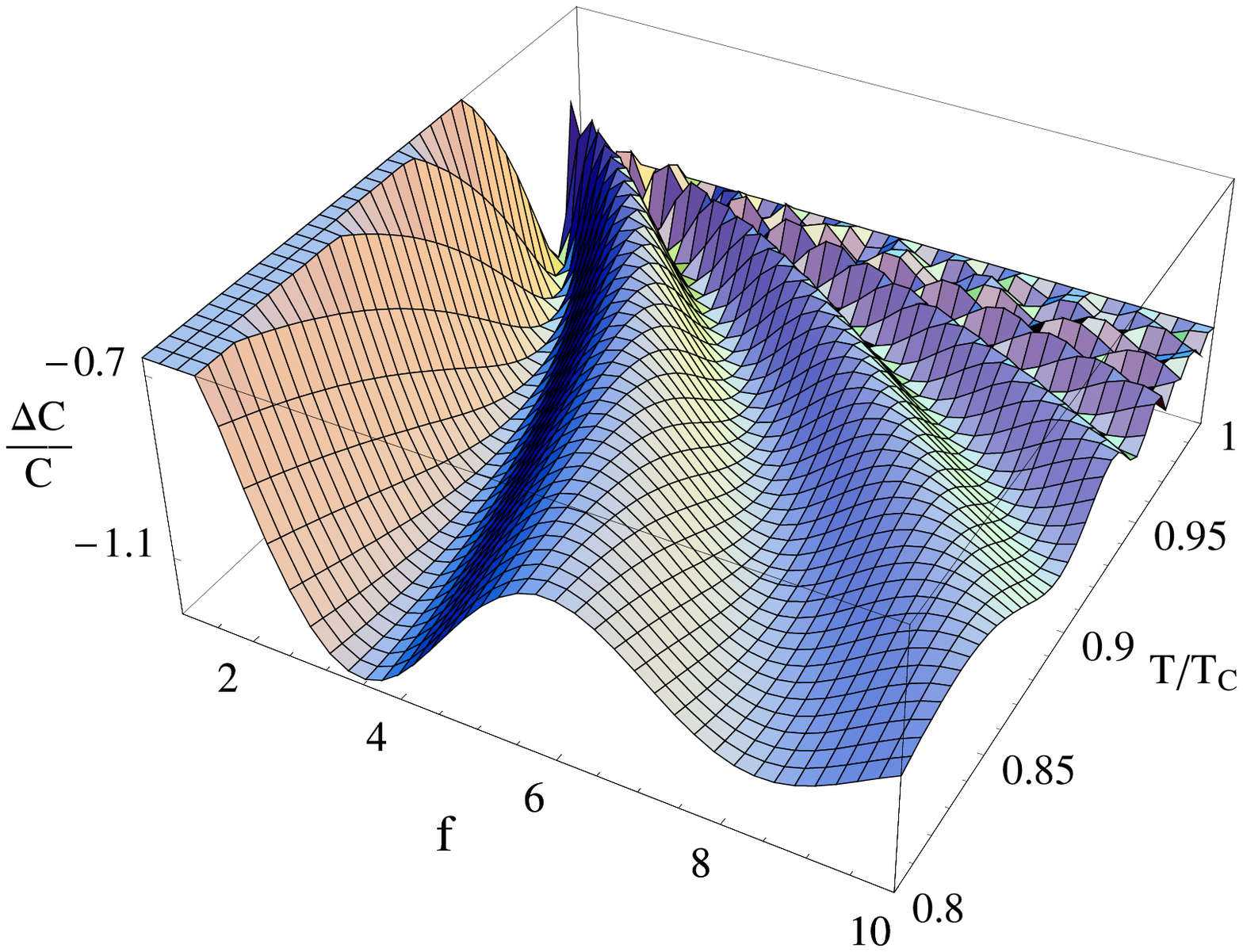}}\vspace{0.25cm}
\caption{ $3D$ flux-temperature  profile of the diaelastic effect
$\Delta C(T,H)/C(T,H)$.} \label{fig:fig4}
\end{figure*}

{\bf 3. Results and Discussion.} Fig.~\ref{fig:fig1} presents the
temperature behavior of the contact area shear modulus $C(T,0)$
(with $t(0)/\xi = 1$, $\xi /\lambda _L(0)=0.02$ and $\beta =0.1$)
in zero applied magnetic field. Notice that $C(T,0)$ is positive
for all temperatures. The considered here field induced analog of
the diaelastic effect means {\it softening} of the contact area
shear modulus under the influence of the applied magnetic field
with $\Delta C(T,H)=C(T,H)-C(T,0)<0$. Fig.~\ref{fig:fig2}
demonstrates this predicted behavior showing the field dependence
of the DE $\Delta C(T,H)/C(T,H)$ for different temperatures. As it
would be expected from the very structure of Eqs.(1)-(9), the DE
of a single contact exhibits {\it field} oscillations imposed by
the Fraunhofer dependence of the critical current $I_C$.  Even
more interesting is its temperature dependence. Indeed, according
to Fig.~\ref{fig:fig3}, depicting the temperature dependence of
the DE for different values of the frustration parameter
$f=H/H_0(0)$, we see characteristic flux driven {\it temperature}
oscillations of $\Delta C(T,H)$ near $T_C$. These oscillations are
governed by the temperature dependence of the London penetration
depth $\lambda_L(T)$, which controls the number of fluxons
entering Josephson contact at a given temperature via the
characteristic field $H_0\simeq \Phi _0/2\pi \lambda _{L}w$. A
more spectacular view of the double temperature-flux oscillations
of the DE can be seen through its $3D$ image, presented in
Fig.~\ref{fig:fig4}. The predicted here effect should manifest
itself not only in single JJs and highly ordered JJAs but also in
granular superconductors (described as disordered JJAs) including
the so-called nanogranular (or intrinsically granular)
superconductors ~\cite{2,4}. In the latter case, however, the
influence of Abrikosov vortices on weak-link mediated DE should be
taken into account due to high values of the characteristic field
$H_0$ reaching a few Tesla for contact size $w$ of a few
nanometers.

In summary, by considering an elastic response of a small
Josephson contact to an effective applied stress field, an analog
of the so-called diaelastic effect was predicted to occur in such
a contact manifesting itself as a magnetic field induced softening
of its shear modulus with pronounced field and temperature
oscillations.

This work has been financially supported by the Brazilian agencies
CAPES, CNPq, and FAPESP.

\end{document}